# ARTICLE

# Realising biaxial reinforcement via orientation-induced anisotropic swelling in graphene-based elastomers†

Mufeng Liu,[a] Ian A. Kinloch,[a] Robert J. Young,[a] Dimitrios G. Papageorgiou.[a,b*]

The biaxial mechanical properties constitute another remarkable advantage of graphene, but their evaluation has been overlooked in polymer nanocomposites. Herein, we provided an innovative and practical method to characterise biaxial reinforcement from graphene via swelling of elastomers, where graphene nanoplatelets (GNPs) were controlled to be oriented in-plane. The in-plane-aligned graphene imposed a biaxial constraining force to the elastomer during the swelling process that led to the anisotropic swelling behaviour of the bulk nanocomposites. This unconventional swelling behaviour was successfully modelled using statistical mechanics. Novel *in situ* Raman band shift measurements were also performed during the de-swelling process of the samples. The characteristic 2D band shift of the GNPs, imposed from stress transfer during the de-swelling procedure, was of the order of 0.1 cm$^{-1}$/% strain, enabling the effective biaxial Young's modulus of the GNPs in the nanocomposites to be calculated (~2 GPa). The determination of the biaxial modulus of GNPs contributed to the evaluation of their Poisson's ratio (~0.34), that is very important but highly impractical to be measured directly on nanoscale specimen.

## 1. Introduction

Graphene has attracted a great deal of attention from industry and academia since its first isolation back in 2004.[1] The high aspect ratio of graphene facilitates the formation of a large interfacial area between the filler and the matrix, which in turn enhances the mechanical properties of polymers.[2] Moreover, the unique two-dimensional geometry of graphene offers biaxial reinforcement in-plane, which is commonly overlooked with regard to its utilization in polymer nanocomposites.

Thermoplastic elastomers (TPEs) have been reinforced by graphene-based fillers to improve their multifunctional properties.[3-5] Most of the studies focus on polyurethane-based plastics, such as thermoplastic polyurethane,[3-5] where graphene enables self-healing of the elastomers and enhances the dynamic mechanical, electrical and thermal properties of the TPE. However, a number of commercially-available polymer blends also show great promise for a number of applications. A member of this family of polymers that was studied herein, is Alcryn™, a thermoplastic vulcanizate based on amorphous polymer blends, with a fraction of the component being chemically crosslinked. Importantly, it is an environmentally-friendly and melt-reprocessible TPE with excellent rubber-like mechanical properties which can be improved further with the introduction of functional fillers such as graphene nanoplatelets.[8]

Elastomers, possessing large free volume, usually display relatively poor liquid barrier properties, allowing small molecules of liquids to diffuse into the materials quite easily.[6] The diffusion of liquid molecules into elastomers results in a uniform three-dimensional swelling of the materials. In this case, for any two principal axes during the swelling (or de-swelling) process, the elastomer can be considered under biaxial deformation. Hence, for an elastomer nanocomposite reinforced by graphene, if the orientation of the two-dimensional nanofiller can be controlled to be unidirectional (or in-plane), it is possible to evaluate the biaxial effective mechanical properties of the graphene in the elastomer nanocomposite.

The transport of small molecules into polymer specimens with different geometries has been investigated by Cervenka and co-workers based on Fickean diffusion.[7] Relative parameters, such as the diffusion coefficient ($D$) and the mass uptake at saturation ($M_\infty$), were specified and well-defined for relatively small specimens. Their specific analysis can be also applied to elastomer-based nanocomposites, in order to investigate the effect of the filler on the absorption of small molecules.

Since the swelling of elastomers in a solvent is related to the mechanical behaviour of the specimen,[8, 9] the mechanism of reinforcement of the filler in an elastomer composite is believed to be through constraining stress provided by the fillers.[9] Raman spectroscopy has been proven to be a powerful technique in characterising interfacial stress transfer from a polymer matrix to graphene-based materials, normally under uniaxial tension.[2, 10-12] Kueseng and Jacob[13] studied a series of elastomer

[a.] *National Graphene Institute and Department of Materials, School of Natural Sciences, The University of Manchester, Oxford Road, Manchester M13 9PL, UK.*
[b.] *School of Engineering and Materials Science, Queen Mary University of London, Mile End Road, London E1 4NS, UK.*

composites reinforced by carbon nanotubes (CNTs) and found that the characteristic Raman bands (G and 2D band) of the CNTs displayed a red shift after the sample was aged in water. This implies that it should be possible to also use Raman spectroscopy to follow the reinforcing mechanisms of graphene during the swelling of an elastomer nanocomposite.

In this work, graphene nanoplatelets (GNPs) were incorporated into a thermoplastic vulcanizate (Alcryn®) by melt mixing. The controlled in-plane orientation of the 2D flakes within the nanocomposites was achieved through compression moulding and the orientation was further verified by scanning electron microscopy and polarised Raman spectroscopy. Additionally, anisotropic swelling was validated by dimensional swelling measurements in toluene. The Flory-Rhener theory and Treloar's expansion of the Flory-Rhener theory[14-16] have been employed to model the anisotropy of swelling. Raman spectroscopy has been employed for the first time to monitor the in-plane, strain-induced 2D band shifts during the de-swelling process, in order to explore the mechanism of reinforcement by GNPs in the elastomer nanocomposites. An estimation of the Poisson's ratio of the GNPs used is given, in agreement with the values obtained from molecular structural mechanics models that were reported.

## 2. Experimental Methods

### 2.1 Materials

Graphene nanoplatelets (GNPs) with nominal lateral diameters of 5, 10, 25 μm and average thicknesses in the range of 6-8 nanometers (according to the supplier) were purchased from XG Sciences, Inc. Lansing, Michigan, USA and used as received. Three grades of xGnP® namely M5, M15 and M25 were used and in a previous study[2] and their diameters were found to be 5.2±3.3 μm, 6.6±4.1 μm and 7.7±4.2 μm respectively. The thermoplastic vulcanizate, Alcryn® 2265 UT (Unfilled Translucent), which is based on a chemically-crosslinked chlorinated olefin interpolymer alloy, was purchased from A. Schulman, Inc.

### 2.2 Preparations

The melt mixing of the composites was carried out in a Thermo Fisher HAAKE Rheomix internal mixer. The mixing took place at 165 °C and 50 rpm for 5 minutes. The GNP fractions in the nanocomposites were 1%, 5%, 10%, and 20% by weight. The Alcryn nanocomposites in this study are coded based on the type of the matrix, the diameter and weight content of the fillers. For example, the sample code 2265-M5-GNP1, means that the matrix is the Alcryn 2265, the diameter of the GNPs is 5 μm and the weight percentage of the GNP filler is 1 wt%.

After the melt mixing, the composite materials were hot pressed into sheets (~2 mm thick) in a Collin platen press (Platenpress P300 P/M). The moulding took place at 185 °C for 10 mins under a hydraulic pressure of 10 bar. The moulded sheets were then stamped into discs with a diameter of ~25 mm for swelling tests and dumbbell samples for tensile tests.

### 2.3 Characterisations

The actual loadings of GNPs in the nanocomposites were obtained by thermogravimetric analysis (TGA) using a TA Q500 TGA instrument. The samples were heated from room temperature up to 600 °C at 10°C/min under a 50 mL/min flow of $N_2$.

Dynamic mechanical analysis (DMA) using a TA Q800 DMA instrument was employed to study the glass transition temperature of the neat polymer and the nanocomposites reinforced by GNPs. The hot-pressed elastomer samples with thickness of 1.5 mm were cut to 40 mm × 5 mm rectangular strips and were tested under tension between -90 °C and 50 °C. The swelling tests were undertaken using toluene (purity of 99.8%, Sigma-Aldrich). The weight, diameter and thickness of the samples were measured after immersion in toluene for 0, 15, 30, 45, 60, 90, 120, 180, 300, 540, 1020, 1980 and 2880 minutes. At least 3 specimens were measured for each sample. The morphologies of the neat polymer and the microstructure of the nanocomposites were examined using scanning electron microscopy (SEM). Both moulded elastomer sheets and swollen and dried samples were evaluated. The cryo-fractured cross-sections of the samples were coated using an Au-Pd alloy in order to provide satisfactory conductivity to the samples. The images were acquired using a high-resolution Tescan Mira 3 Field Emission Gun Scanning Electron Microscope (FEGSEM) at 5 kV.

Tensile testing was undertaken using an Instron 4301 machine with a load cell of 5 kN, for all the samples. At least 5 specimens were tested for each sample. The measurements were undertaken at a tensile rate of 50 mm·min$^{-1}$ in accordance with ASTM 638.

Fourier transform infrared (FTIR) was employed to carry out the compositional analysis of the materials. The spectra were obtained using a NICOLET 5700 spectrometer.

Polarised Raman spectroscopy was employed to evaluate the spatial orientation of the filler in the matrix. A similar method was used in our previous work.[10, 11, 17, 18] The equipment was a 514 nm Raman spectrometer by Renishaw with 'VV' (vertical-vertical) polarization, in which the incident and scattered radiation were polarised in the same direction. The laser was aligned perpendicular to the surface of the materials either along the axis of in-plane or cross-plane. The intensity of the Raman G bands was recorded as a function of the rotation angle to enable estimation of the orientation distribution function (ODF).[17]

Raman spectroscopy was also employed to measure the in-plane strain-induced 2D band shifts. Raman spectra were acquired using a Renishaw InVia Raman spectrometer with a laser wavelength of 633 nm and an objective of 50×, which produces a spot size of 1–2 μm. Ideally, it would have been best to follow the band shifts of the GNPs in the nanocomposites during swelling in toluene. In practice, this proved to be difficult and it was decided to follow the band shifts during the de-swelling of swollen specimens during natural evaporation of the solvent. The Raman 2D band shifts of the swollen nanocomposite samples (after 3 hours immersion in toluene) with the highest loading of GNPs (20 wt%) was first measured. The in-plane strain, determined by measuring the diameter of the disc-shape samples with a digital caliper, reduced with time of exposure the air and the change in band position was determined. The Raman laser spot was focused on the same point of a single flake on each sample surface. The results were based on 3 composite samples for each type of GNP, at the highest loading. All spectra were fitted with a single Lorentzian curve.

## 3. Results and Discussion

### 3.1 Microstructure and orientation of the filler

The microstructure of the materials was characterised by SEM as shown in Fig. 1. It can be seen that the GNPs are well-dispersed and well-oriented (Fig. 1, a to c) in the elastomer as a result of the preparation procedure for loadings up to 20 wt%. The GNPs appear to be well wetted by the elastomer as suggested by the high magnification images in Fig. 1 (d to f) and so the interface between the matrix and the flakes can be considered satisfactory. The preparation of graphene-reinforced elastomer nanocomposites by melt mixing followed by injection/compression moulding, usually leads to the in-plane orientation of the filler in the matrix due to the shear and compression involved.[10, 11] During the compression moulding process employed in this work, the compression force in the cross-plane $z$ direction (shown in the Fig. S1a) gave rise to a preferred in-plane orientation ($x$-$y$ plane) of the flakes (shown in Fig. S1, b and c). The preferred in-plane alignment of the filler leads to anisotropic reinforcement, where the mechanical Polarised Raman spectroscopy was employed to quantify the degree of orientation the graphene nanoplatelets in the matrix. The specific method was also utilized in a number of our previous studies upon GNP-reinforced polymers.[10, 11, 17, 18] The dependence of the Raman intensity on the angle of incident laser in both in-plane and cross-plane is shown in the Fig. 1 (g and h). When the Raman laser was perpendicular to the cross-plane direction, the intensity of the G band remained unchanged with varying polarization angles. However, when the polarised laser was parallel to the in-plane direction, the angular dependence of the Raman intensity of the G band on the polarization angle is quite different. In this case the G band Raman intensity decreases from 0° to 90° (and 180° to 270°) and then increases from 90° to 180° (and 270° to 360°). The quantification of the in-plane flake orientation can be accomplished by curve fitting using the following Equation for the in-plane direction data:[17]

$$I_{\text{sample}}(\Phi) = I_o \left\{ \frac{8}{15} + \langle P_2(\cos\theta) \rangle (-\frac{16}{21} + \frac{8}{7}\cos^2\Phi) + \langle P_4(\cos\theta) \rangle (\frac{8}{35} - \frac{8}{7}\cos^2\Phi + \cos^4\Phi) \right\} \quad (1)$$

where $I_o$ is the amplitude and assuming the surface normals are uniformly distributed around the cross-plane axis. properties of the samples in the in-plane direction ($x$-$y$ plane) are improved.

This phenomenon is also reflected in the swelling experiments performed in this work and will be discussed in detail in the next section.

The orientation factor for the in-plane reinforcement is then given by:

$$\eta_o = \frac{8}{15} + \frac{8}{21} \langle P_2(\cos\theta) \rangle + \frac{3}{35} \langle P_4(\cos\theta) \rangle \quad (2)$$

The orientation factors for the nanocomposite samples at the highest filler loading (20 wt%) were quantified: $\eta_{oM5}$=0.89, $\eta_{oM15}$=0.84 and $\eta_{oM25}$=0.80. Generally, $\eta_o$ should be between 0.53 (random orientation) and 1 (perfect in-plane orientation). Therefore, it can be concluded that the GNPs are well oriented in the in-plane direction in the elastomer sheets.

### 3.2 Diffusion of the solvent

The elastomeric matrix and the nanocomposite specimen were immersed in toluene and measurements of the swollen specimen were performed in order to investigate the relative mass uptake $M$, and the diffusion coefficient $D$. For well-defined, small, plate-like specimens (with a diameter of at least 10× the thickness), the relative mass uptake $M(t)$ determined gravimetrically with exposure times is given by:[7]

$$M(t) = \frac{W(t) - W(0)}{W(0)} \quad (3)$$

where $W(t)$ is the weight of a specimen after an exposure time $t$ and $W(0)$ is calculated theoretically by determining the intercept of the linear fitting of $W(t)$ against $t^{1/2}$ at $t$=0, in order to reduce any systematic error involved in gravimetry.[7]

The diffusion coefficient, $D$, can then be determined by:[7]

$$\frac{4M(\infty)}{h\sqrt{\pi}}\sqrt{D} = \frac{M_2 - M_1}{\sqrt{t_2} - \sqrt{t_1}} \quad (4)$$

where $h$ is the thickness of the samples and $M(\infty)$ is the mass uptake at the saturation point of the absorption.

The relative mass uptake $M(t)$ as a function of $t^{1/2}$ for the elastomer composites reinforced by three types of GNPs is shown in Fig. 2 (a to c). At the beginning of exposure time, the small molecules of the solvent were absorbed into the materials leading to an increase of $M(t)$. Afterwards, the curves form a plateau at $t^{1/2} \approx 100$ (s$^{1/2}$), which indicates the saturation of the absorption phenomenon. The elastomer used in this study is based on a thermoplastic vulcanizate, where a fraction of the macromolecular chains are chemically crosslinked.[19] Hence, the relative mass uptake $M$(t) is found to decrease slightly after reaching the maximum $M$(t) values at $t^{1/2} \approx 100$, due probably to slow dissolution of the polymer, shown in Fig. S2. In this work, the $M(\infty)$ values are defined for consistency as the maximum M(t) values for all measurements.

The relative mass uptake $M(t)$ as a function of $t^{1/2}$ for the elastomer composites reinforced by three types of GNPs is shown in Fig. 2 (a to c). At the beginning of exposure time, the small molecules of the solvent were absorbed into the materials leading to an increase of $M(t)$. Afterwards, the curves form a plateau at $t^{1/2} \approx 100$ $(s^{1/2})$, which indicates the saturation of the absorption phenomenon.

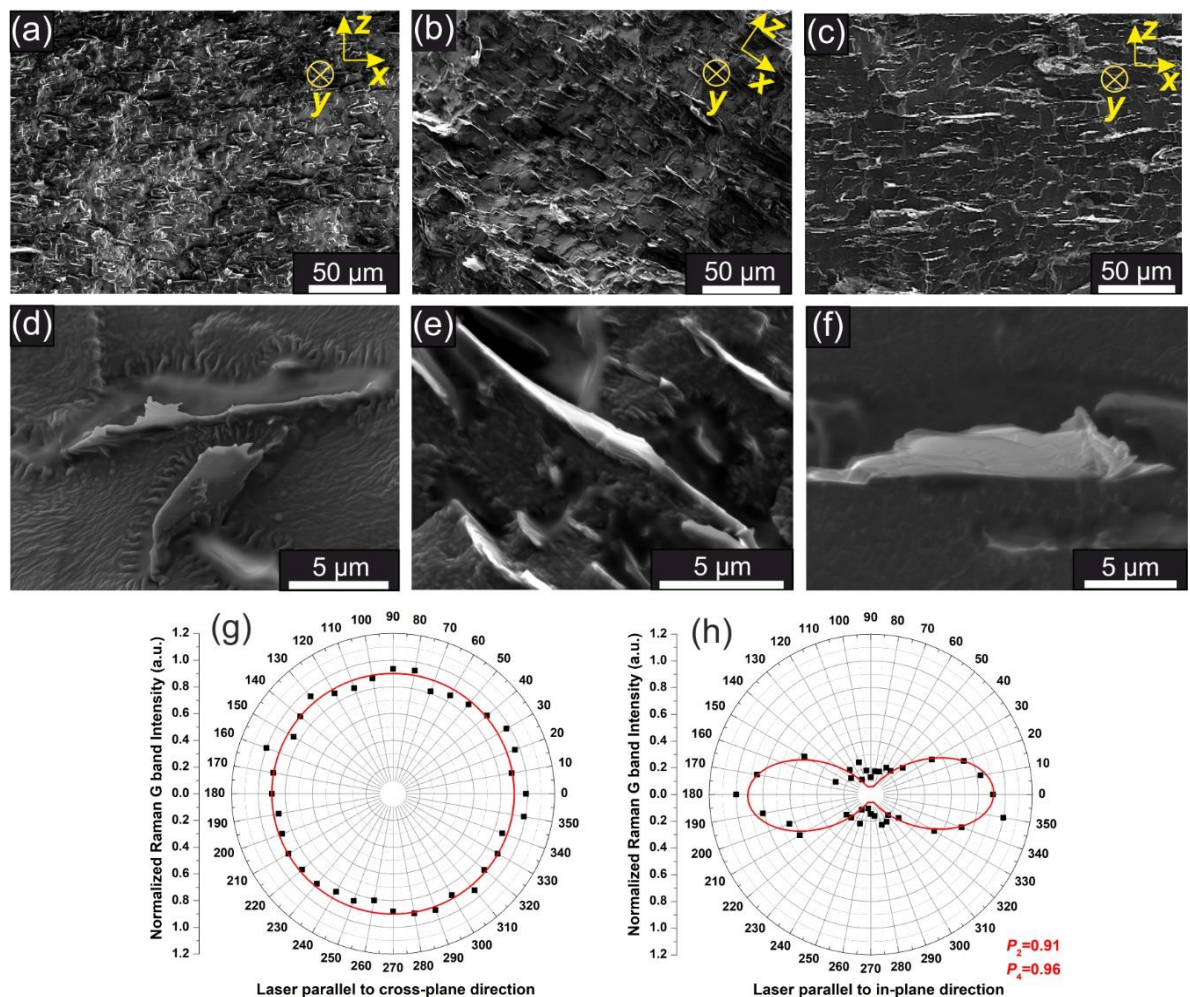

**Fig. 1** Microstructure of the nanocomposites and filler orientation. (**a-c**) Low magnification SEM images of cryo-fractured cross-sectional surface of the elastomer nanocomposite sheets (20 wt% of M5, M15 and M25 GNP-reinforced nanocomposites, respectively) showing good dispersion of the filler; (**d-e**) High magnification SEM images of individual M5, M15 and M25 GNP flakes embedded in the nanocomposites showing good wetting. The dependence of normalized Raman G band intensity on the angle of the incident laser parallel to (**g**) the cross-plane direction and (**h**) the in-plane direction, measured for an M5 GNP-reinforced TPE at the highest loading (20 wt%).

The elastomer used in this study is based on a thermoplastic vulcanizate, where a fraction of the macromolecular chains are chemically crosslinked.[16] Hence, the relative mass uptake $M(t)$ is found to decrease slightly after reaching the maximum $M(t)$ values at $t^{1/2} \approx 100$, due probably to slow dissolution of the polymer, shown in Fig. S2. In this work, the $M(\infty)$ values are defined for consistency as the maximum $M(t)$ values for all measurements.

From Fig. 2 (a to c), it can be seen that $M(\infty)$ is significantly reduced with increasing filler content, exhibiting an effective restriction of the ultimate solvent uptake with the introduction of GNPs. When a neat polymer is exposed in a solvent, the small molecules of the solvent can diffuse into the free volume, making the polymer swell in three dimensions, until equilibrium between the stress in the reinforcement and the volume swelling is achieved. When a filler is incorporated into the matrix, the internal stress caused by the swelling of the matrix can be transferred from the matrix to the filler at the filler/matrix interface, which allows the filler to partially carry the load and eventually restrain the overall mass uptake of the materials.

The dimensions of the samples at each time interval were also recorded during the tests and are plotted in Fig. S3 for all specimens. In addition, the geometric swelling ratios of the samples at the saturation point are plotted in Fig. 2 (d to f). As expected, the volume swelling ($V_s/V_0$) of the composite specimen is restrained by the presence of GNPs, similar to the mass uptake. In addition, it can be seen from Fig. 2 (d to f) that the diameter swelling ratio ($d_s/d_0$) of the samples (in-plane) was significantly reduced with increasing GNP content. It is interesting, however, to find that the thickness swelling ratio ($h_s/h_0$) of the samples (cross-plane) increased with increasing filler content. Such results imply that the flakes are highly oriented in the in-plane direction (as was also concluded from the SEM micrographs and the polarised Raman experiments) so that the stress transfer from the matrix to the filler takes place more efficiently in the in-plane direction of the elastomer sheet. Nevertheless, the GNP-filled nanocomposites displayed a higher level of swelling in thickness than the unfilled elastomer, indicating the absence of reinforcement from the GNPs in the out-of-plane direction. The anisotropic swelling behaviour of the filled samples will be analysed in the next section with the application of the Flory-Rhener theory.

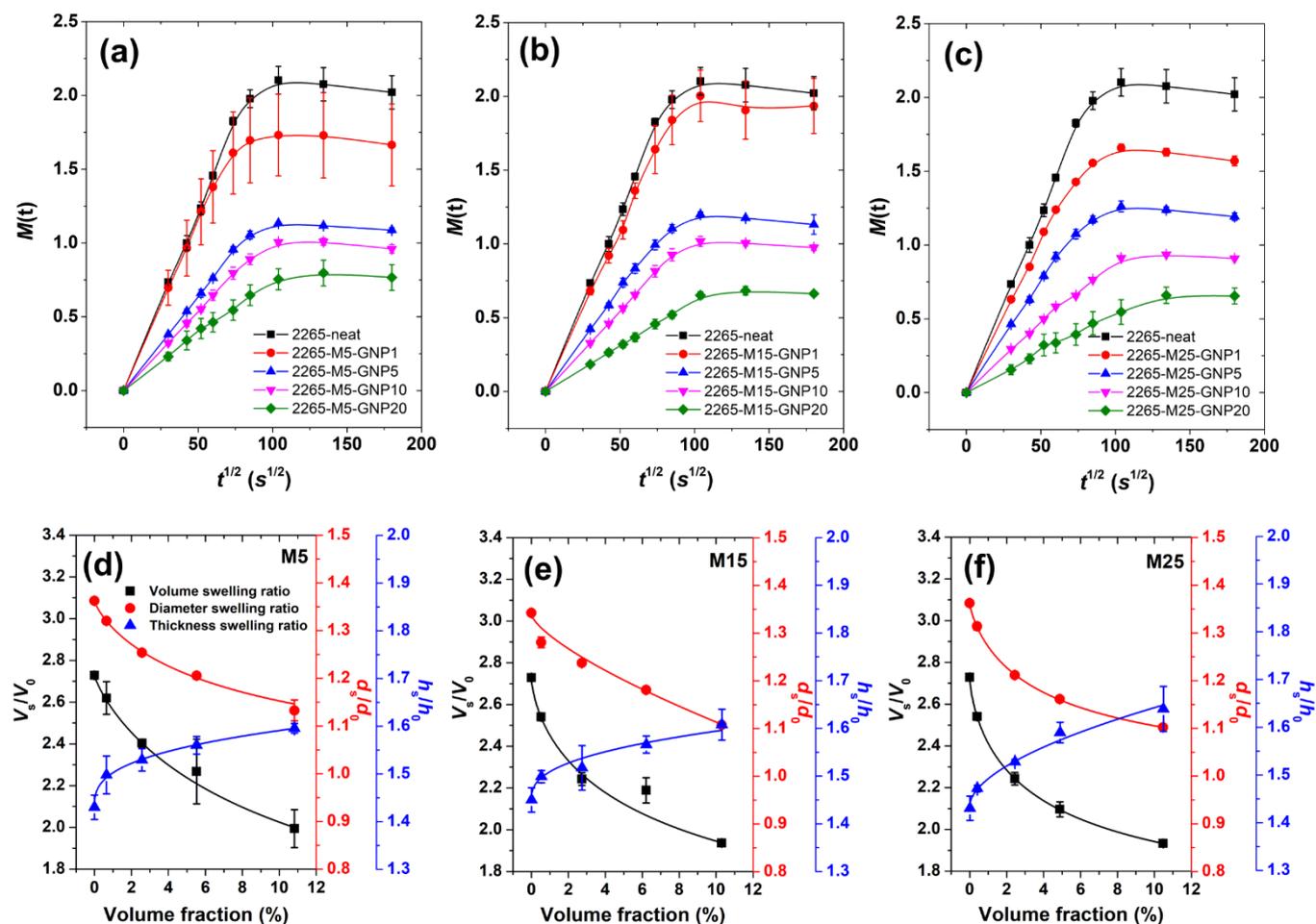

**Fig. 2** Results of mass uptake and geometric swelling. (**a-c**) The mass uptake against the square root of time for the samples of M5, M15 and M25 reinforced elastomers, respectively. (**d-f**) Dependence of the swelling ratio at saturation of volume, diameter and thickness of the samples on filler volume fraction for M5, M15 and M25 GNP reinforced elastomers. In the figure, $V_s/V_0$ is the volume swelling ratio, $d_s/d_0$ is the diameter swelling ratio and $h_s/h_0$ is the thickness swelling ratio.

The diffusion coefficient ($D$) (Table 1) indicates the rate of solvent diffusion. It is highest for the neat elastomer and decreases with increasing loadings for each size of GNPs. It can also be seen that for a given GNP loading it decreases with increasing GNP diameter, going from M5 to M15 to M25 for both the 10 wt% and 20 wt% GNP-reinforced nanocomposites. This behaviour is the result of nanoconfinement of the macromolecular chains of TPE due to the presence of GNPs that can be also confirmed by dynamic mechanical analysis (DMA) (Fig. S4). It can be clearly seen in Fig. S4 that the glass transition temperature ($T_g$) increases with increasing filler loading as suggested by the shift of the major peak of the loss factor curves towards higher temperatures. This indicates a decrease of the mobility of the TPE macromolecules with increasing temperature and filler content (results also listed in Table S1).

### 3.3 Anisotropy of swelling

Anisotropic swelling has been observed previously for fibre-reinforced elastomers or gels and the Flory-Rhener theory has been applied to evaluate the experimental results.[20-22] Similarly to our study, anisotropic swelling was reported in a unidirectionally-aligned, lamellar-structured heterogeneous hydrogel, where the in-plane swelling ratio was lower than cross-plane swelling ratio.[20] In this work, the in-plane swelling ratio ($d_s/d_0$) and cross-plane swelling ratio ($h_s/h_0$) have been derived (see S5 for the detailed derivation of the two equations) for all samples under study, based on the Flory-Rhener equation[14-16] and are given by the following equations:

$$d_s/d_0 = (1/v_{2\ \text{neat}})^{-5/12} \cdot (1/v_2)^{3/4} \quad (5)$$

$$h_s/h_0 = (1/v_{2\ \text{neat}})^{5/6} \cdot (1/v_2)^{-1/2} \quad (6)$$

where the $d_s$ and $h_s$ are the diameter and thickness of the swollen samples at saturation point, $d_0$ and $h_0$ are the diameter and thickness of the unswollen samples. The parameter $v_2$ is defined as the volume fraction of polymer in the swollen specimen and so $1/v_2$ is the swelling ratio.[6, 15, 16] The volume swelling ratio at saturation point is defined as $1/v_{2\,neat}$ for neat polymer and $1/v_2$ for the biaxially reinforced composites.

Table 1. Results of diffusion coefficient. The volume fraction determined by TGA and the diffusion coefficient of toluene into the nanocomposites evaluated by the swelling tests.

| 2265-M5 | | 2265-M15 | | 2265-M25 | |
|---|---|---|---|---|---|
| GNP loading (vol%) | $D$ (×$10^{-5}$) (mm$^2$/s) | GNP loading (vol%) | $D$ (×$10^{-5}$) (mm$^2$/s) | GNP loading (vol%) | $D$ (×$10^{-5}$) (mm$^2$/s) |
| 0 | 9.89 ± 1.21 | 0 | 9.89 ±1.21 | 0 | 9.89 ± 1.21 |
| 0.65 | 9.40 ± 0.45 | 0.53 | 9.22 ± 0.13 | 0.40 | 9.60 ± 1.14 |
| 2.57 | 9.12 ± 0.26 | 2.73 | 9.17 ± 0.07 | 2.44 | 9.06 ± 0.02 |
| 5.52 | 8.42 ± 1.13 | 6.21 | 7.66 ± 0.16 | 4.88 | 7.62 ± 0.01 |
| 10.82 | 7.66 ± 0.27 | 10.3 | 6.78 ± 0.05 | 10.47 | 5.95 ± 1.03 |

In equations (5) and (6), the fixed quantities $(1/v_{2\,neat})^{-5/12}$ and $(1/v_{2\,neat})^{5/6}$ introduce the intrinsic properties of the polymer/solvent system. As mentioned earlier, the theoretical basis for these equations is the well-known Flory-Rhener equation:[16] $\rho V_1/M_c \approx (1/2-\chi)(1/v_{2\,(neat)})^{-5/3}$, where the experimentally-obtained swelling ratio of the neat polymer at saturation ($1/v_{2\,neat}$) can be correlated with the properties of the polymer (density $\rho$, molecular weight of chain $M_c$), the solvent (molar volume $V_1$) and the polymer-solvent interaction (dimensionless parameter $\chi$). The variables $(1/v_2)^{3/4}$ and $(1/v_2)^{-1/2}$ in equations (5) and (6), on the other hand, represent the dependence of the in-plane and cross-plane swelling ratios ($d_s/d_0$ and $h_s/h_0$ respectively) on the swelling ratio of the volume $(1/v_2)$ after the introduction of in-plane-aligned GNPs into the elastomer. Equation (5) indicates that the diameter swelling ratio ($d_s/d_0$) decreases with decreasing volume swelling ratio $(1/v_2)$, while equation (6) suggests that the thickness swelling ratio ($h_s/h_0$) increases with decreasing volume swelling ratio $(1/v_2)$. With the introduction of GNPs, the volume swelling ratio $(1/v_2)$ of the nanocomposites decreases due to interfacial stress transfer to the GNPs. This will subsequently lead to an increase of the cross-plane thickness swelling ratio ($h_s/h_0$) based on equation (6), which also explains why the thickness swelling ratio of the nanocomposites is even greater than that of the neat polymer as shown in Fig. 2 (d to f). It is critically important to point out that the Flory-Rhener theory was originally developed for the study of the swelling of rubbers, rather than nanocomposites. Therefore, one of the basic conditions for applying the Flory-Rhener equation for the calculation of the crosslinking density is that the rubber is deformed isotropically during the swelling process.[14-16] A number of studies[24-29] have applied an analogy of "physical crosslinks" that actually result from the constraining stress that normally builds up at the filler/matrix interface. As a consequence, the overall crosslinking density that was calculated using the original FR theory[24-29] increased with increasing filler loadings, since the final reduction of the swelling ratio comprised the contribution of both the chemical and "physical" crosslinking. In the present study, anisotropic swelling was clearly exhibited and the phenomenon can be attributed to the in-plane 'constraining tension' from the GNPs; thereby the original Flory-Rhener can only be used for qualitative estimation. Treloar's model was introduced here to analyse the swelling behaviours of the elastomer under biaxial constraining force, as discussed in detail in S5 (Supplementary Information).

In Fig. 3, the black and red curves are the plots of equations (5) and (6), using the experimentally-measured volume swelling ratio of the neat polymer ($1/v_{2\,neat}$). Theoretically, the point of intersection of the two curves at $1/v_2 \approx 2.75$ defines isotropic swelling. However, a slight difference was observed between the experimentally-obtained diameter and the thickness swelling ratios of the neat elastomer, as a result of in-plane orientation of the polymer chains, originating from the compression moulding procedure. Moreover, as can be seen in Fig. 3, the data points of the measured swelling ratios in diameter and thickness of the samples fit extremely well to the curves for derived equations (5) and (6). From the intersection point towards lower swelling ratio of the volume $(1/v_2)$, the difference between the swelling ratio of diameter ($d_s/d_0$) and thickness ($h_s/h_0$) becomes larger as the filler loading increases, indicating higher degree of biaxial mechanical constraint in-plane from the GNPs. The efficiency of biaxial reinforcement of larger flakes is again better than that of smaller flakes.

It should be emphasized that chemical components or bonds remained unchanged prior to and after the polymer were melt mixed with GNPs, as can be confirmed from the FTIR spectra of the nanocomposites, shown in Fig. S6. The elastomer has a chlorinated component that enables the materials to be melt processable and ameliorates the interpolymer miscibility. The anisotropic swelling resulted from the anisotropic modulus of the nanocomposite samples due to the in-plane filler orientation. Thereby, we also evaluated the mechanical properties (in S7). The modulus of the materials in the in-plane direction was increased significantly with increasing loading due to effective stress transfer.[11] In the out-of-plane direction, however, the modulus of the nanocomposites can be considered similar to that of the matrix. We established a correlation between the modulus and the dimensional swelling ratios in S8, showing good agreement between the results and the theory, indicating once again the validity of the newly-produced equations (5) and (6) in Fig.3.

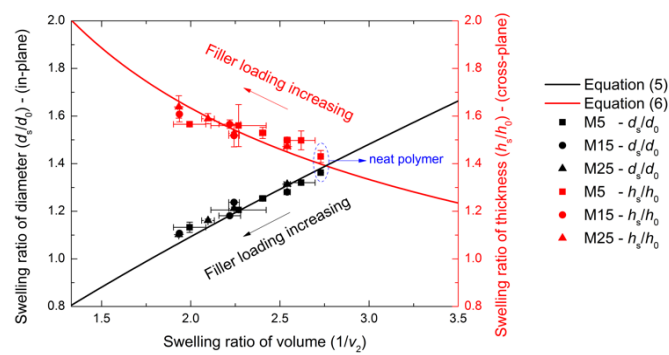

**Fig. 3** Application of derived anisotropic swelling theory on experimental results. The dependence of swelling ratio of diameter ($d_s/d_0$) and thickness ($h_s/h_0$) on the swelling ratio of volume ($1/v_2$). The red curve and black curve are plotted based on equations (5) and (6). The datapoints were measured experimentally.

### 3.4 Biaxial deformation-induced Raman band shifts

Since in-plane reinforcement by the graphene nanoplatelets was realised from the swelling measurements, a further study of the stress transfer was carried out by evaluating the Raman band shifts during de-swelling. When a swollen specimen is taken out from the solvent and exposed in the air, the solvent molecules tend to evaporate from it. This causes the strain in the GNPs to reduce and consequently the internal stress at the filler/matrix interface also decreases. Such a process allowed us to monitor the blue shift of the characteristic 2D Raman band of the GNPs, as shown in Fig. 4(a). In this experiment, only the nanocomposites with the highest filler loading (20 wt% of the GNP) were examined, since the high filler loading made the volatilization of the solvent slow enough to allow the Raman spectra to be captured. The Raman spectra of the GNPs and the diameters of the composite specimens were simultaneously recorded. The peak positions of the 2D band are plotted against strain in Fig. 4 (b to d). It can be seen that the Raman 2D band shifts to higher wavenumbers with the decrease of the in-plane strain in the nanocomposites. Generally, larger flakes give higher 2D band shift per strain, which is in agreement with the diffusion and swelling results reported previously. The reason for this is that for short platelets embedded in a matrix, stress transfer takes place at the edge of the flakes and increases from the edges towards the centre, along the nanoplatelets.[2] The data in Fig. 4 (b to d) indicate the better reinforcement by the larger nanoplatelets in M15 and M25 showing increasingly-higher values of -d$\omega_{2D}$/d$\varepsilon$, and therefore higher effective biaxial modulus (Fig. 4e).

The assessment of the in-plane reinforcement of the GNPs can be achieved by analysing the band shift values listed in Table 2. The effective modulus of graphene in the elastomer matrix under biaxial deformation can be calculated by:[2]

$$E_{\text{eff}} = -\frac{d\omega_{2D}}{d\varepsilon} \cdot \frac{E_{\text{biaxial}}}{(\partial\omega_{2D}/\partial\varepsilon)_{\text{biaxial-ref}}} \text{GPa} \quad (7)$$

where -d$\omega_{2D}$/d$\varepsilon$ is the 2D Raman band shift per composite strain, $E_{\text{biaxial}}$ is the biaxial modulus of monolayer graphene measured by pressurized graphene "balloons", which is equal to 2400 GPa as reported by Lee et al.[24] and $(\partial\omega_{2D}/\partial\varepsilon)_{\text{biaxial-ref}}$ is the 2D Raman band shift rate of monolayer graphene under controlled biaxial deformation, which is equal to -148 cm$^{-1}$/% strain, according to the work of Androulidakis et al.[25] The effective modulus values of the GNPs in the elastomer matrix for this work were calculated using equation (7) and are listed in Table 2. The theoretical effective modulus of the filler measured by Raman band shifts ($E_R$) and based on the shear-lag theory for uniaxial deformation is given by:[2]

$$E_{R\ \text{uniaxial}} \approx \eta_o \frac{s^2}{8} \frac{1}{1+v_{\text{matrix}}} \frac{t}{T} E_m \quad (8)$$

where $\eta_o$ is the orientation factor of the filler, $s$ is the aspect ratio of the measured flake, $v_{\text{matrix}}$ is the Poisson's ratio of the matrix $t$ is the thickness of the flake and $T$ is the thickness of the polymer layer surrounding the flake in the model nanocomposite and $E_m$ is the modulus of the matrix. It should be pointed out that Equation (8) shows that the effective modulus of the filler depends principally upon the Young's modulus of the matrix, $E_m$, the nanoplatelet aspect ratio, $s$ and thickness, $t$.

For the biaxial deformation in this work, the modulus of the matrix is given by $E_m/(1-v_{\text{matrix}})$, since the stresses ($\sigma_x$ and $\sigma_y$) and strains ($\varepsilon_x$ and $\varepsilon_y$) along the principal axes are equal. In addition, both the nanoplatelets and the matrix were under biaxial deformation during the de-swelling process. Hence, the theoretical effective biaxial modulus of the filler measured by Raman band shift is given by:

$$E_{R\ \text{biaxial}} = \eta_o \frac{s^2}{8} \frac{1}{1+v_{\text{matrix}}} \frac{E_m}{1-v_{\text{matrix}}} \frac{1}{1-v_{\text{graphene}}} \frac{t}{T} \quad (9)$$

The Poisson's ratio of elastomers is generally considered to be approaching 0.5.[26] The Poisson's ratio of graphitic materials is a matter of debate as it ranges widely in the literature from 0.16 to 0.43,[27-36] mainly on the basis of theoretical predictions. Hence, it can be inferred theoretically that $E_{R\ \text{biaxial}}/E_{R\ \text{uniaxial}}$ should be in the range of 2.4 to 3.5. It can be seen (Table 2) that the biaxial deformation gives an effective modulus of the GNPs close to the one that was expected, based on the ratio $E_{R\ \text{biaxial}}/E_{R\ \text{uniaxial}}$ (average values around 3) for all three types of GNPs (listed in Table 2). From equation (9) we are able to calculate the Poisson's ratio of the graphene nanoplatelets, which ranges

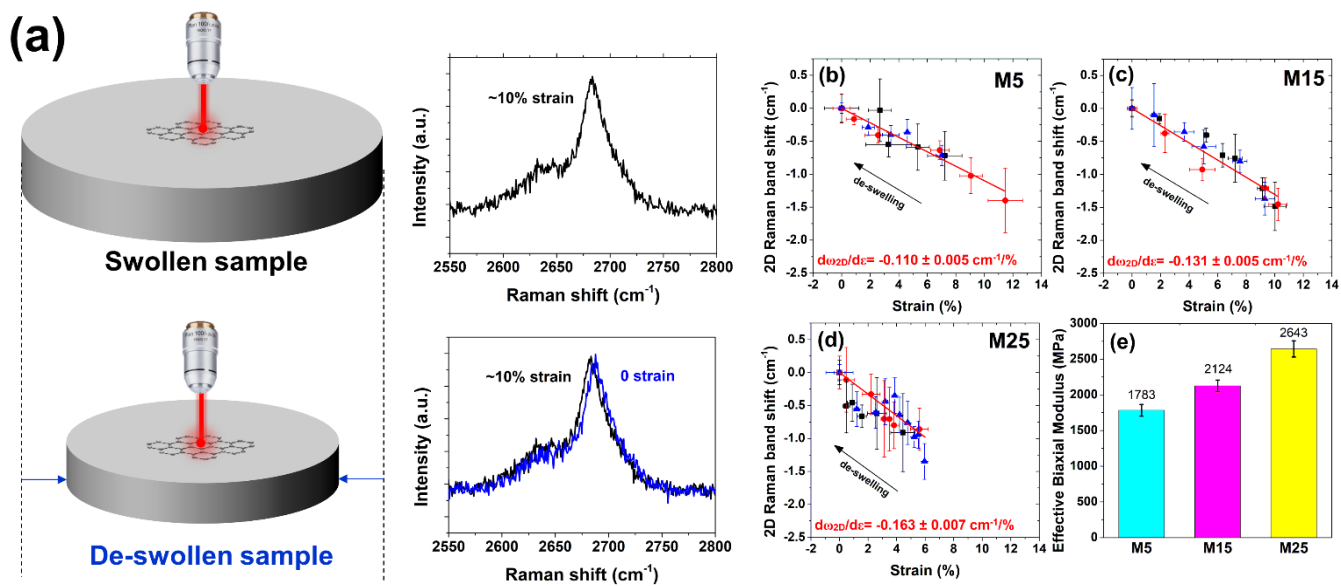

**Fig. 4** Raman band shift during de-swelling. (**a**) Schematic diagram of the measurement of the 2D Raman band shift with an example of spectra at different in-plane strains. The in-plane strain here is given by $\varepsilon_{in\text{-}plane}= (d_{swollen}–d_{dry})/d_{dry}$, where $d_{swollen}$ is the diameter of the swollen samples, and $d_{dry}$ is the diameter of the dried samples after immersion in the toluene and dried in the air rather than the original diameter before immersed in the solvent. (**b-d**) Raman 2D band shift of M5, M15 and M25 GNPs as a function of the corresponding strains during the swelling, respectively. The continuous lines are linear fits to the experimental data. (**e**) Effective biaxial modulus of M5, M15 and M25 GNP flakes calculated from Raman 2D band shift (**b to d**) using equation (7).

**Table 2.** Calculated properties from Raman 2D band shift. Raman 2D band shift and the calculated effective modulus for M5, M15 and M25 GNPs under both uniaxial and biaxial deformation are listed, where $d\omega_{2D}/d\varepsilon$ is the Raman 2D band shift per strain and $E_R$ is the effective modulus of the GNPs measured by Raman band shift and calculated by the equation (7).

|  | M5 | M15 | M25 |
|---|---|---|---|
| Diameter (μm) [2] | 5.2±3.3 | 6.6±4.1 | 7.7±4.2 |
| $d\omega_{2D}/d\varepsilon$ (cm$^{-1}$/%)(uniaxial) [11] | -0.036±0.003 | -0.040±0.005 | -0.050±0.006 |
| $E_R$ (MPa)(uniaxial) [11] | 630±53 | 700±88 | 875±105 |
| $d\omega_{2D}/d\varepsilon$ (cm$^{-1}$/%)(biaxial) | -0.110±0.005 | -0.131±0.005 | -0.163±0.007 |
| $E_R$ (MPa)(biaxial) | 1783±81 | 2124±81 | 2643±114 |
| $E_R$ (bi)/ $E_R$ (uni) | 2.99±0.50 | 3.10±0.50 | 3.09±0.50 |
| Average Poisson's Ratio | 0.33 | 0.35 | 0.35 |

from 0.33 to 0.35.

This result is also consistent with the Poisson's ratio of carbon nanotubes (0.34), determined using the local density approximation model.[37] Furthermore, Zhao et al.[28] computed the Poisson's ratio of graphene to be around 0.30, by utilizing a molecular structural mechanics model. The accurate value could only be realised experimentally by nanoscale measurements, determining the axial and transverse strain simultaneously, which is very hard to achieve. Hereby, we have provided a rather accessible route to measure the Poisson's ratio of the graphene substances indirectly by introducing micromechanical theories that can be employed in polymer nanocomposites along with the use of in-situ Raman spectroscopy.

## 4. Conclusions

Swelling measurements of the nanocomposite samples in an organic solvent (toluene) indicated that both the diffusion of the liquid molecules and the swelling of the elastomers are significantly restrained by the introduction of GNPs. This is the first report of anisotropic reinforcement from 2D materials in elastomers as a result of swelling. The reinforcement from the nanoplatelets during the swelling of the nanocomposites is highly dependent upon the orientation of the GNP flakes. The extended theoretical analysis based on the Flory-Rhener theory has shown a very good agreement with the experimentally-observed anisotropic swelling of the samples. The study of Raman 2D band shifts during the de-swelling process has demonstrated directly that the mechanism of biaxial reinforcement is interfacial stress transfer from the matrix to the GNPs. Additionally, from the use of Raman spectroscopy on biaxially deformed nanocomposites, we were able to provide an indirect experimental method for the estimation of the Poisson's ratio of graphene, which is of high significance for future studies.

Excellent liquid barrier properties are imperative to elastomers, due to their wide applications in a liquid environment including seals, gaskets, hoses/bonded flexible pipes and joints, valve sleeves and others. From the analysis of experimental results, we have shown that GNPs exhibit a high reinforcing efficiency to suppress the swelling of elastomers and the orientation of the fillers is of vital importance for the control of swelling in different directions. This work can act as a guideline for the evaluation of the mechanisms of biaxial reinforcement, the factors affecting functional performance and the prediction of the modes of deterioration in swollen GNP-elastomer nanocomposites, so that end users can be given confidence to continue to use nanocomposite elastomeric components often to completion of their design life. As has been demonstrated, the introduction of graphene-related materials into elastomers can bring great benefits and enhanced properties for a number of practical applications.

## Conflicts of interest

There are no conflicts to declare.

## Acknowledgements

This project has received funding from the European Union's Horizon 2020 research and innovation programme under grant agreement No 785219. Ian A. Kinloch also acknowledges the Royal Academy of Engineering and Morgan Advanced Materials.